\documentclass[sigconf]{acmart}

\AtBeginDocument{%
  \providecommand\BibTeX{{%
    \normalfont B\kern-0.5em{\scshape i\kern-0.25em b}\kern-0.8em\TeX}}}

\setcopyright{none}
\copyrightyear{2024}
\acmYear{2024}
\acmDOI{XXXXXXX.XXXXXXX}

\usepackage{xspace,xcolor}
\usepackage{multirow, multicol} 
\usepackage{todonotes}
\usepackage{listings}
\usepackage{caption}
\usepackage{subcaption}
\usepackage{pgf-pie}

\usepackage{CJKutf8}
\usepackage{enumitem}
\setlist{leftmargin=5.5mm}

\newcommand{\codefuse}{\textsc{CodeFuse}\xspace} 
\newcommand{\codefusellama}{\textsc{CodeFuse-CodeLlama}\xspace} 
\newcommand{\codegeex}{\textsc{CodeGeeX}\xspace} 
\newcommand{\codellama}{\textsc{CodeLlama}\xspace} 
\newcommand{\codegen}{\textsc{CodeGen}\xspace} 
\newcommand{\starcoder}{\textsc{StarCoder}\xspace} 
\newcommand{\ernie}{\textsc{BaiduErnie}\xspace} 
\newcommand{\neox}{\textsc{GPT-NeoX}\xspace} 
\newcommand{\humaneval}{\textsc{HumanEval}\xspace} 
\newcommand{\mbpp}{\textsc{MBPP}\xspace} 
\newcommand{\mbxp}{\textsc{MBXP}\xspace} 
\newcommand{\codefuseeval}{\textsc{CodefuseEval}\xspace} 
\newcommand{\company}{AntGroup\xspace} 

\newcommand{\tu}[1]{{\small\texttt{#1}}}

\definecolor{mauve}{rgb}{0.88, 0.69, 1.0}

\usepackage{wasysym}

\copyrightyear{2024}
\acmYear{2024}
\setcopyright{acmlicensed}\acmConference[ICSE-SEIP '24]{46th International Conference on Software Engineering: Software Engineering in Practice}{April 14--20, 2024}{Lisbon, Portugal}
\acmBooktitle{46th International Conference on Software Engineering: Software Engineering in Practice (ICSE-SEIP '24), April 14--20, 2024, Lisbon, Portugal}
\acmDOI{10.1145/3639477.3639719}
\acmISBN{979-8-4007-0501-4/24/04}

\begin{document}

\title{CodeFuse-13B: A Pretrained Multi-lingual Code Large Language Model}



\author{Peng Di$^\dag$, Jianguo Li$^\dag$, Hang Yu$^\dag$, Wei Jiang$^\dag$, Wenting Cai, Yang Cao, Chaoyu Chen, Dajun Chen, Hongwei Chen, Liang Chen, Gang Fan, Jie Gong, Zi Gong, Wen Hu, Tingting Guo, Zhichao Lei, Ting Li, Zheng Li, Ming Liang, Cong Liao, Bingchang Liu, Jiachen Liu, Zhiwei Liu, Shaojun Lu, Min Shen, Guangpei Wang, Huan Wang, Zhi Wang, Zhaogui Xu, Jiawei Yang, Qing Ye, Gehao Zhang, Yu Zhang, Zelin Zhao, Xunjin Zheng, Hailian Zhou, Lifu Zhu, Xianying Zhu}
\authornote{Non-corresponding authors are listed in alphabetical order.}
\affiliation{%
  \institution{Ant Group, China}
  \country{$^\dag$Corresponding-authors: \{dipeng.dp,lijg.zero,hyu.hugo,jonny.jw\}@antgroup.com}
}

\renewcommand{\shortauthors}{Peng Di, et al.}

\begin{abstract}
Code Large Language Models (Code LLMs) have gained significant attention in the industry due to their wide applications in the full lifecycle of software engineering. However, the effectiveness of existing models in understanding non-English inputs for multi-lingual code-related tasks is still far from well studied.
This paper introduces \codefuse-13B, an open-sourced pre-trained code LLM \footref{foot:link}. It is specifically designed for code-related tasks with both English and Chinese prompts and supports over 40 programming languages. \codefuse achieves its effectiveness by utilizing a high-quality pre-training dataset that is carefully filtered by program analyzers and optimized during the training process.
%
Extensive experiments are conducted using real-world usage scenarios, the industry-standard benchmark \humaneval-x, and the specially designed \codefuseeval for Chinese prompts. To assess the effectiveness of \codefuse, we actively collected valuable human feedback from the \company's software development process where \codefuse has been successfully deployed. 
The results demonstrate that \codefuse-13B achieves a \humaneval pass@1 score of 37.10\%, positioning it as one of the top multi-lingual code LLMs with similar parameter sizes. In practical scenarios, such as code generation, code translation, code comments, and testcase generation, \codefuse performs better than other models when confronted with Chinese prompts.




\end{abstract}

\ccsdesc[500]{Software and its engineering}

\keywords{code large language models, multi-lingual, Chinese prompts}



\maketitle





\section{Introduction}

Code Large Language Models (Code LLMs) have attracted substantial attention in the industry owing to their vast applications throughout the entire software engineering lifecycle.
The release of Copilot, empowered by Codex~\cite{chen2021evaluating}, served as a significant testament to the imminent arrival of the era of intelligent code.
One astonishing application, ChatGPT~\cite{brown2020language,openai2023gpt4}, has captivated an incredible user base of over 100 million in two months since its launch. 
In recent code models such as AlphaCode\cite{AlphaCode}, InCoder\cite{incoder},
SantaCoder\cite{santacoder}, StarCoder\cite{starcoder}, and Code Llama\cite{codellama}, the incorporation of fill-in-the-middle capabilities has proven to be particularly valuable for practical code completion scenarios.

While these models have practical applications to software development processes, their effectiveness in comprehending non-English inputs for code-related tasks remains unsatisfactory\cite{zhao2023survey}. To bridge this gap, \codegeex \cite{codegeex} attempted to establish a connection between code and non-English languages, incorporating vocabulary tokens from various natural languages. Indeed, by leveraging large, domain-specific datasets, LLMs can significantly enhance their effectiveness in applications that necessitate a combination of natural language understanding and domain-specific knowledge, including specialized terminology.


\begin{table}[]
  \caption{\codefuse  project roadmap.}
  \label{tab:timeline}
  \vspace{-1ex}
  \begin{tabular}{clc}
    \toprule
    {\textbf{Release}} & \multirow{2}{*}{\textbf{Model}} & \textbf{\humaneval} \\
     \textbf{date}&& \textbf{Pass@1} \\
    \midrule
    Mar 2023&\codefuse-1.3B-2K Seq-Length&11.58\%\\
    Apr 2023&\codefuse-6.5B-4K Seq-Length&20.46\%\\
    May 2023&\codefuse-13B-Base-4K Seq-Length&32.93\%\\
    Jun 2023&\codefuse-13B (opened in Sep) & 37.10\%\\
    Sep 2023&\codefusellama-34B (opened)& 74.40\%\\
  \bottomrule
\end{tabular}
\vspace{-1ex}
\end{table}

The paper introduces \codefuse,
a Language Model for coding,  which is open-sourced on GitHub~\footnote{\url{https://github.com/codefuse-ai}} and Huggingface~\footnote{\url{https://huggingface.co/codefuse-ai}\label{foot:link}}, an open-source code Language Model (LLM). The \codefuse project, a collaborative effort within \company, has witnessed monthly model iterations resulting in consistent performance improvements, as depicted in Table~\ref{tab:timeline}. As of this September, we open-sourced two versions of \codefuse, \codefuse-13B and \codefusellama-34B. \codefuse-13B underwent fine-tuning by LoRA/QLoRA on multiple code tasks using the self-pretrained base model, while \codefusellama-34B was fine-tuned using CodeLlama-34b-Python. Excitingly, \emph{\codefuse-13B surpasses other code LLMs of similar size, and \codefusellama-34B outperforms GPT4 and ChatGPT-3.5 on the \humaneval benchmark.}

This paper centers around the pre-trained model \codefuse-13B, providing a comprehensive overview of the development process and evaluating its performance in real industrial scenarios. The production of \codefuse-13B encompasses several crucial steps, including:
\begin{itemize}
\item Data collection: We collected about 200+ TB of code-related data, and finally refined it to around 1.6TB (1T Token) of clean data suitable for pre-training.

\item Program feature analysis: We extracted a set of program features from the collected code, including syntax correctness, cleanliness score, etc. This analysis serves three purposes: 1) ensuring high-quality code for pre-training data, 2) providing AST/CFG/DFG/IR program feature data and extracting code semantics to facilitate program understanding, and 3) profiling the code dataset to guide constraint-based instruction fine-tuning. The analyzer employed a datalog-based program analysis method, translating analysis into a query system. 

\item Pre-training: Using the \company's common technology stack, we developed \codefuse to pre-train a large-scale model with 13 billion parameters in a stable manner.

\item Instructional fine-tuning: This stage involved various fine-tuning techniques, such as supervised instruction fine-tuning (SFT), and multi-task instruction fine-tuning (MFT), among others.

\item Model evaluation: We provide a comprehensive evaluation kit that supports both online and offline inference evaluation methods. This kit drives model training across different downstream tasks based on performance indicators and visualizes feedback on evaluation results, enabling continuous iterations and optimization of corresponding data, algorithms, and engineering challenges.

\item Model operations: In the context of hundreds or even thousands of GPU training instances, numerous challenges emerge, such as timely automatic detection of faulty nodes, automatic initiation of training tasks, and monitoring the convergence of training progress. We have made significant advancements in addressing these challenges through Model Operations.


\end{itemize}

We conducted an industry-based evaluation of \codefuse by integrating it into the software development process at \company. Additionally, we developed extensions for several popular Integrated Development Environments (IDEs), namely VSCode, JetBrains, and Ant CloudIDE (a Web IDE).
Moreover, we developed and open-sourced a more comprehensive benchmark, named \codefuseeval
\footnote{\label{foot:codefuseeval} \url{https://github.com/codefuse-ai/codefuse-evaluation}.}
, to support for a broader range of programming scenarios involving Chinese inputs.

The results demonstrate that \codefuse-13B achieves a \humaneval Pass@1 score of 37.10\%, making it one of the top models with similar sizes. In practical multi-lingual scenarios, \codefuse outperforms other models with Chinese prompts, such as code translation, code comments, testcase generation and others.



We summarize the key contributions as follows:

\begin{itemize}
    \item We introduce \codefuse-13B, an open-sourced pre-trained code LLM with 13B parameters and its training procedure. It is specifically designed for code-related tasks with Chinese prompts and supports over 40 programming languages. 
    \item We have developed \codefuse extensions for various IDEs. These extensions enable developers to seamlessly integrate \codefuse into their coding workflows, enhancing productivity and code generation capabilities.
    \item We evaluate the effectiveness of \codefuse in various application scenarios, including code generation, code translation, code comments, and testcase generation. The results demonstrate that \codefuse-13B achieves a \humaneval Pass@1 score of 37.10\%, outperforming other multi-lingual models with similar sizes. Moreover, \codefuse-13B is better than other models in practical scenarios involving Chinese prompts.
\end{itemize}


\section{Data Preparation}

\subsection{Overview of data processing}
\begin{figure}[]
\usetikzlibrary{shadows,arrows}

\tikzstyle{practica} = [rectangle, text width=10em, minimum width=12em, minimum height=3em, draw=black, thick,inner sep=1pt, text centered, rounded corners]

\tikzstyle{moreinfo} = [rectangle, text width=10em, minimum width=12em, minimum height=3em, draw=black, thick,inner sep=1pt, text centered, rounded corners, dashed]

\tikzstyle{line} = [draw, thick, color=black!80, -latex']

\newcommand{\practica}[2]{node (#1) [practica]
  {#1\\{\scriptsize{#2}}}}

\centering
\begin{tikzpicture}[scale=0.9,transform shape,->,>=stealth']
  \path \practica {Data acquisition}{\parbox{16em}{200T raw data including 196T code, 1.75T Chinese and 1.7T English.
  }};
  \path (Data acquisition.east)+(2.5,0.0) \practica{Data cleaning}{ 
  \parbox{16em}{a) File attribute filtering\\b) Code quality filtering}};

  \path (Data cleaning.south)+(0.0,-1.0) \practica{File attribute filter}{
  \parbox{16em}{
    Filter by file-level rules including file size, code proportion, etc
   }};
  \path (File attribute filter.south)+(0.0,-1.0) \practica{Code quality filter}{
  \parbox{15em}{
    a) Syntax verification\\
    b) code feature analysis
   }};
  \path (File attribute filter.west)+(-2.5,0.0) \practica{Data detoxification}{
  \parbox{15em}{\company's content risk control
and privacy data protecting}};
  \path (Data detoxification.south)+(0.0,-1.5) \practica{Data deduplication}{
\parbox{15em}{a) MD5 file-level deduplication\\
b) SimHash fine-grained file-level deduplication\\
c) Segment-level deduplication
}};
  \path (Data deduplication.south)+(0.0,-1.0) \practica{Data resampling}{
    \parbox{16em}{
Resampling some languages for the model's training effectiveness}};

  \path [line] (Data acquisition.east) -- node [left] {} (Data cleaning);
  \path [line] (Data cleaning.south) -- node [above] {} (File attribute filter);
  \path [line] (File attribute filter.south) -- node [above] {} (Code quality filter);
  \path [line] (Code quality filter.west) -- node [right] {} (Data detoxification);

  \path [line] (Data detoxification.south) -- node [above] {} (Data deduplication);
  \path [line] (Data deduplication.south) -- node [above] {} (Data resampling);  

\end{tikzpicture}
\caption{The diagram of data processing.}
\label{fig:code_process}
\end{figure}
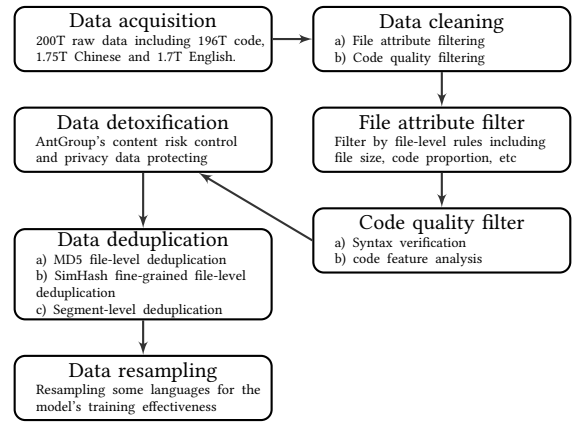

Since code LLMs focus more on code-related tasks, there are significant differences compared to the methods used in general LLMs, 
in terms of data acquisition, data cleaning, detoxification, deduplication, and data resampling, as shown in Figure~\ref{fig:code_process}. This paper will primarily focus on how \codefuse constructs pre-training data for the code domain large model.

\begin{figure}[t]
\centering
    \begin{tikzpicture}[scale=.9, font=\scriptsize]
\pie{14.0/Java,
    12.3/Python,
    11.7/C++,
    10.5/JavaScript,
    9.8/C,
    6.4/Markdown,
    5.6/PHP,
    5.2/TypeScript,
    5.2/C\#,
    4.3/Go,
    3.6/HTML,
    1.2/CSS,
    1.1/Objective-C,
    9.1/Others}
 
\end{tikzpicture}
\caption{Distribution of programming languages.}
\label{fig:code_dis}
\end{figure}
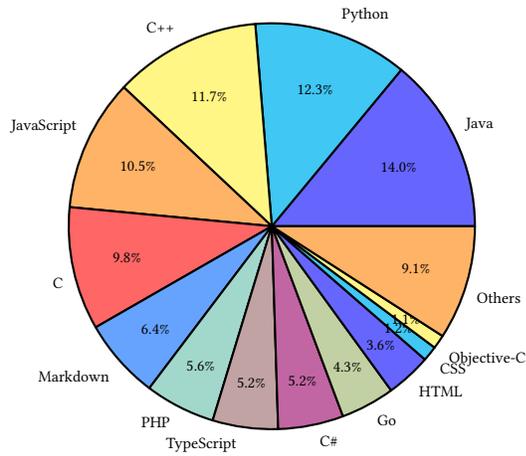

\textbf{Data acquisition. }
The pre-training data for \codefuse consists of 196TB of code, 1.75TB of Chinese raw data, and 1.7TB of English raw data, totaling 200TB, that are tokenized into
800 billion tokens of code, 100 billion tokens of Chinese corpus, and 100 billion tokens of English corpus (see Section~\ref{sec:token}).
The original code data were a combination of self-crawled GitHub dataset and opensource dataset \tu{Stack}~\cite{kocetkov2022stack}.
The combined dataset is deduplicated and filtered to include over 40 programming languages with the distribution shown in Figure~\ref{fig:code_dis}, in which 13 programming languages account for more than 1\% each. Java, Python, C++, and JavaScript surpass 10\%, while the remaining nearly 30 programming languages comprise 9\% of the total.
The Chinese corpus is sourced from \tu{CommonCrawl}, computer-related websites, documentation of programming languages and their third-party libraries, etc. The English corpus is sampled from various categories in \tu{Pile} including \tu{StackExchange}, \tu{Arxiv}, \tu{Wikipedia}, \tu{OpenWebText2}, \tu{Pile-CC}, etc.

\textbf{Data cleaning. }
The cleaning strategy for code data is divided into two levels. The first-level filtering strategy involves the aspect of file attributes, including strategies such as discarding large files (e.g., files with more than 10,000 lines or single files larger than 1MB) and discarding abnormal text (e.g., lines with an average length greater than 100 or a proportion of alphanumeric characters less than 40\%).
The second-level filtering strategy relies on \company's program analyses~\cite{sparrow, pinpoint, pengseip22a, pengseip23a,pengissta23} to filter out code data that does not meet the requirements of syntax correctness and code quality. The open-source release of our program feature analyzer is planned, and the details will be introduced in Section~\ref{sec:programanalysis}.



\textbf{Data detoxification.}
We utilize \company's content risk control and privacy data protecting capabilities to identify and filter out risky data from the training dataset, ensuring data safety.

\textbf{Data deduplication. }
Multiple granularities of deduplication are performed to sensuality of training data.
First is the global deduplication with file-level \tu{MD5}.
The second is fine-grained file-level 
deduplication 
with \tu{SimHash} score (e.g. $\ge$0.95).
Third is segment-level deduplication based on code analysis to separate codes and comments, and deduplicate code and text segments when document-level \tu{SimHash} score is larger than a threshold (e.g. $\ge$0.90). 

\textbf{Data resampling. }
Language distribution-based resampling removes data for niche programming languages (with a data proportion below 0.1\%) and downsamples HTML, CSS, JSON, and other similar languages that can negatively impact the model's training effectiveness.

Through these data processing stages, the training data for the \codefuse large model is prepared to ensure high-quality data for training and enable the model to exhibit excellent performance.
The filtered and resampled data are tokenized into a format directly usable for the pre-training phase.



\subsection{Program feature analysis}\label{sec:programanalysis}



To improve the quality of training code data, we propose two approaches to analyze code features: static analysis-based and model-based methods. Static analysis-based method offers cost-effectiveness, interpretability, and iterative refinement but may struggle with quantifying complex code features. The model-based method provides better handling of unquantifiable features but comes with higher costs and weaker interpretability. For massive code data, static analysis is preferred first, yet a hybrid approach combining static analysis with model-based classification can be used for fine-grained cleansing. This approach leverages the strengths of both methodologies to ensure efficient and effective handling of code features.

\textbf{High-quality code evaluation model. }
To effectively evaluate the quality of code, we have proposed several metrics integrated into our model. These metrics serve as indicators of code quality, and they are designed to provide comprehensive insights into potential areas of improvement.
\begin{itemize}
    \item \textbf{Correctness measurement:} After performing syntax verification and bug detection, it has been observed that the number of bugs is inversely proportional to the quality of the code. A higher bug frequency indicates lower code quality. To provide a more detailed analysis, bugs have been categorized into three levels: Fatal, Error, and Warning.
    
    \item \textbf{Readability:} Larger methods, classes, and more method parameters tend to be associated with lower code quality. This is due to the increased complexity and reduced readability that often accompanies large methods, classes, and more parameters.
    \item \textbf{Redundancy:} The presence of redundant classes is a sign of poor code quality. Redundancy in classes often leads to increased complexity and decreased maintainability.
    \item \textbf{Naming style:} Identifier names that are either too long or too short can compromise code quality. Short names may not adequately describe the purpose of a variable or function, while excessively long names can hinder readability.
    \item \textbf{Cyclomatic complexity: }The cyclomatic complexity is a metric used to measure the complexity of the control flow of a module. It quantifies the number of independent paths through the code, which can also be understood as the minimum number of test cases required to cover all possible scenarios.
    \item \textbf{Coupling:} It is a measure of the degree of association between modules. The strength of coupling depends on the complexity of the interfaces between modules, the way modules are invoked, and the amount of data transmitted through interfaces. The coupling between modules refers to the dependency relationship between them, including control relationships, invocation relationships, and data transmission relationships.
\end{itemize}

\textbf{Implementation. } We utilized \company's static analyzer, named \tu{Sparrow}, to filter the code. Similar to \tu{CodeQL}~\cite{codeql} developed by \tu{GitHub}, \tu{Sparrow} is a datalog-based program analysis tool that translates analysis into a query system. To handle the analysis of a substantial volume of code, we stored the resulting information offline in a database along with a datalog solver engine. This setup allows us to efficiently query and retrieve program analysis results.

\subsection{Code semantic extraction}\label{sec:semantic}
One of the most crucial tasks for code LLMs is to comprehend the semantics of the code. To support that, it is needed to extract codes with natural language annotations, which serve as interpretations of the code's semantics. 
Therefore, we employ static analysis to extract code snippets and their corresponding comments from high-quality code. This extracted data is then utilized for the Supervised Fine-tune (SFT) of \codefuse. The goal of this approach is to ensure that the model not only understands the syntax of the programming languages but also gains a deep comprehension of the underlying logic and functionality of the code, thereby enhancing its code understanding capabilities.

\textbf{Strategy for selecting code-comment pairs. }
To ensure the quality of code-comment pairs, we focus on extracting functions and their corresponding comments from the code. We utilize rule-based approaches to filter code-comment pairs: 
\begin{itemize}
    \item \textbf{Meaningless comments:} Meaningless comments are detected by \tu{Sparrow} using a set of rules, which include comment keyword detection, identification of auto-generated setter/getter methods, and others. These comments are considered irrelevant and are subsequently discarded.
    \item \textbf{Code length limits:} Code containing fewer than 3 lines is discarded to ensure sufficient context in the code.
    \item \textbf{Effective code limits:} Methods with less than 60\% of effective code lines are discarded to ensure a significant proportion of meaningful code.
    \item \textbf{Comment length limits:} To maintain readability and manageability, comments longer than 512 characters are discarded.
\end{itemize}

\textbf{Implementation. }
We utilize tree-sitter, a fast and robust parsing tool, to support multiple programming languages and handle large code volumes. By leveraging the tree-sitter's query functionality, we extract function-comment pairs across different languages, enabling a unified data structure. The tool efficiently distributes and aggregates code parsing results, enhancing parsing speed and compatibility with various programming languages.




\textbf{Application. }
We have accumulated a dataset of 3.2 billion function-level code-comment pairs. This dataset spans various sources, including GitHub, GeeksforGeeks, \starcoder, and \company's internal code repositories, providing a broad and diverse base for training and evaluation. The dataset is used on the following code-related tasks (see Section~\ref{sec:eval}):
\begin{itemize}
    \item \textbf{Code generation:} This task involves the translation of natural language into code snippets. The large dataset aids in understanding the semantic meaning behind the natural language and generating the corresponding code.

    \item \textbf{Code comment:} This task focuses on the generation of comments for given code. Leveraging the code-comment pairs in our dataset, \codefuse can effectively generate meaningful and context-specific comments for any given piece of code.

    \item \textbf{Code explanation:} This task deals with providing comprehensive explanations for given code snippets. The vast amount of code-comment pairs within our dataset aids in drawing parallels and generating detailed, understandable explanations of code functionality.
\end{itemize}


\subsection{Code dataset portrait} 

Code dataset portrait refers to an automated approach that involves annotating, categorizing, and analyzing code in order to capture various dimensions of large-scale codebases. By performing code profiling, a more thorough and detailed understanding of the code data used for training can be obtained. This understanding facilitates more efficient implementation of techniques such as Supervised Self-Training (SST) and Self-Supervised Fine-Tuning (SFT).
The applications of code portrait can be categorized as follows:
\begin{itemize}
    \item \textbf{Deep insight of training data: } For each portrait dimension, the portrait analysis can provide measurements like validation loss, perplexity, and even feedback errors. These could show us which portion of data are more difficult to learn. 
    \item \textbf{Augmenting training data: } The portrait feedback urges us
 to augment training data by improving the quantity or quality of the corresponding weak performed dimension.
 On the one hand, since the distribution of training data could significantly affect a model's behavior~\cite{zhou2023devil},
 code portrait allows us to improve the quantity or portion of weak performed dimension according to evaluation feedback. 
 On the other hand, it may urge us to check the quality of the weak performed portion of data and find potential ways to further improve the data quality. 
 Both ways may optimize the model in subsequent experiments.
\end{itemize}



The implementation of code portrait is based on mentioned \tu{Sparrow}. 
Based on the code portrait analysis of the code added to \company's repositories in the past three months, the following key insights have been derived:
\begin{itemize}
    \item Java is the dominant language, accounting for approximately 40.7\% of the newly added code. Within the Java codebase, the majority (over 90\%) is attributed to the SOFA (Scalable Open Financial Architecture) projects. Therefore, the primary emphasis for code generation should be on Java applications related to the SOFA stack.
    
    \item Certain modules within the Java code, namely 'test', 'model', 'facade', and 'dal', have a higher representation. These sections should be regarded as high-priority areas for code generation, given their significance within the codebase. Focusing on these modules will help ensure that the generated code adequately addresses the needs and requirements of the system.

    \item JavaScript constitutes 9.23\% of the newly added code, making it the second most widely used language. Within the JavaScript codebase, the Bigfish framework stands out as the most predominant. Therefore, generating code specifically for the Bigfish framework should be given priority to cater to the significant usage of this framework within the JavaScript codebase.
\end{itemize}



\section{Training}



In this section, we introduce the pre-training procedure of \codefuse including tokenization, model architecture and training procedure.

\subsection{Tokenization \label{sec:token}}

\begin{figure}
\centering
\small
\includegraphics[width=.5\textwidth]{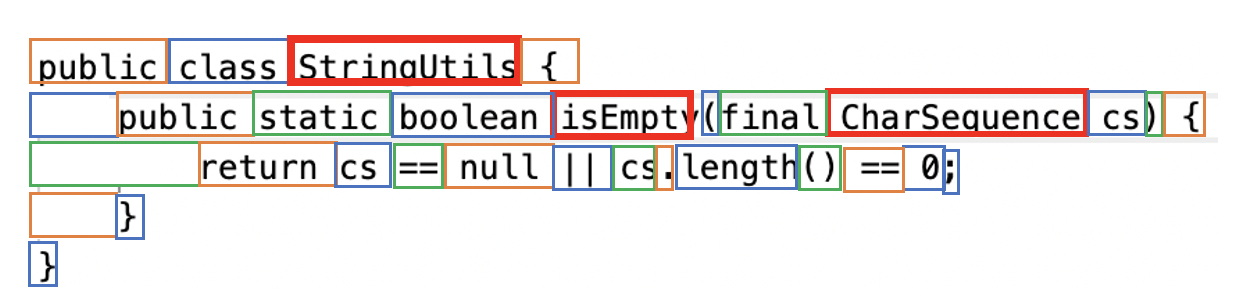}
        \caption*{(a) \codefuse has 40 tokens}
    \includegraphics[width=.5\textwidth]{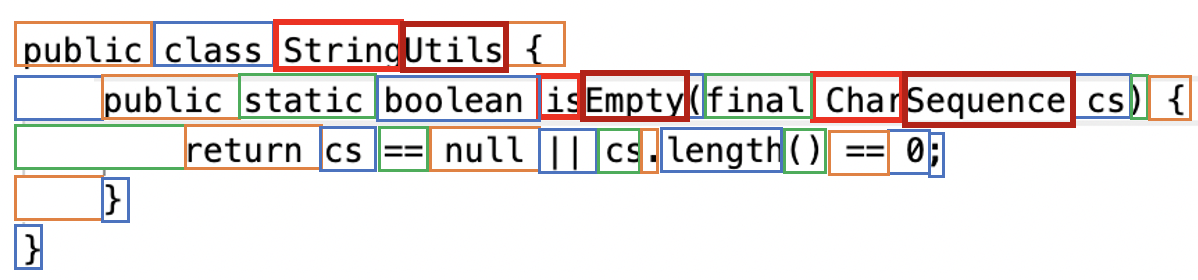}
        \caption*{(b) \codellama has 46 tokens}
    \includegraphics[width=.5\textwidth]{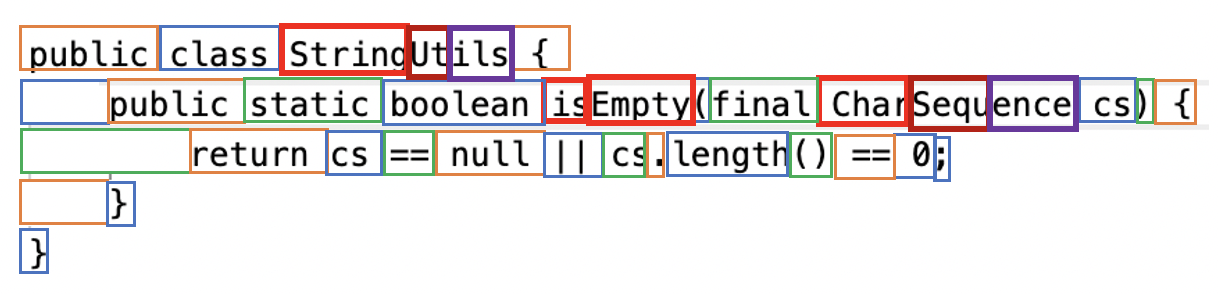}
        \caption*{(c) \codegen has 47 tokens}
\caption{Tokenization examples of different models applied to the same code snippet.}
\label{fig:token-examples}
\end{figure}

Our tokenizer is BPE-based~\cite{Shibata99} and designs to avoid Out of Vocabulary (OOV) problems. It has a vocabulary size of 100,864, covering keywords and common words in programming languages as well as information techniques related corpus.  
The tokenizer also supports effective segmentation of natural languages like Chinese and English. It ensures accurate code tokenization. Figure~\ref{fig:token-examples} provides example cases to demonstrate tokenization by several well-known code LLMs, highlighting the effectiveness of \codefuse's tokenizer.

To verify the tokenization performance of the \codefuse tokenizer on code, Chinese, and English, we randomly sampled 100,000 examples from the training dataset. The token count and compression ratio after tokenization are shown in Table~\ref{tab:token}. The tokenization performance of the \codefuse tokenizer is significantly better than that of \codellama and \codegen for code, Chinese, and English. This is attributed to its larger vocabulary size and dedicated code-specific vocabulary.

\begin{table*}[t]
  \caption{Comparison of compression rate (C-Rate) of tokenization. C-Rate = \#Tokens / \#Characters, the lower the better.}
  \label{tab:token}
  \begin{tabular}{cccccccc}
    \toprule
    \multirow{2}*{\textbf{Type}} & \multirow{2}*{\textbf{\#Characters}} & \multicolumn{2}{c}{\textbf{\codefuse}} & \multicolumn{2}{c}{\textbf{\codellama}}& \multicolumn{2}{c}{\textbf{\codegen}}\\ \cmidrule{3-4} \cmidrule{5-6} \cmidrule{7-8}
    && \textbf{\#Tokens} & \textbf{C-Rate}& \textbf{\#Tokens} & \textbf{C-Rate}& \textbf{\#Tokens} & \textbf{C-Rate}\\
    \midrule
    Code & 338,758,753 & 86,787,734 & 0.25 &99,180,237 &	0.29	& 96,289,455 & 0.28 \\
    Chinese	& 85,998,939  &	98,491,170 &	1.14 &	121,180,842	& 1.41 &	161,977,211 &	1.88 \\
    English	& 283,983,202 &	69,951,060&	0.24&	78,472,584	&0.27&	71,393,619	&0.25 \\
  \bottomrule
\end{tabular}
\end{table*}

\subsection{Model architecture}

\begin{figure}[t]
\centering
\includegraphics[width=.5\textwidth]{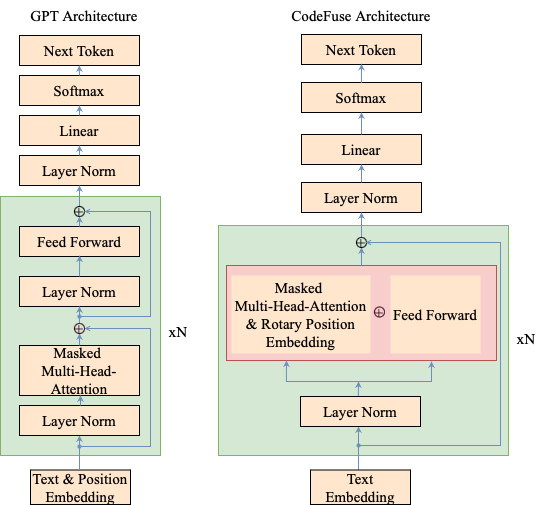}
        
\caption{Comparison  \codefuse architecture with GPT.}
\label{fig:model_arch}
\end{figure}

The model architecture of \codefuse is an auto-regressive Transformer, similar to GPT-3~\cite{brown2020language}, with regular next-token prediction as the learning objective. We made two key changes similar to GPT-J~\cite{gpt-j}: (1) using Rotary positional embeddings instead of learned positional embeddings, and (2) employing parallel attention and feed-forward layers (FFN) instead of serial layers as in GPT3. The details are shown in Figure~\ref{fig:model_arch}.

\textbf{Rotary positional embeddings (RoPE).} We adopt RoPE~\cite{su2022roformer} as position embedding in Multi-Head Attention instead of the learned positional embedding. To balance effectiveness and computational efficiency, we apply RoPE only to the first half of embedding vectors~\cite{black2022gptneox20b, gpt-j}.

\textbf{Parallel attention and FFN layers.} In GPT-3, Multi-Head Attention is computed first, and then the result is added with the residual connection before being passed into FFN as shown in Equation~\ref{eq:ffn1}. In our model, Multi-Head Attention and FFN are computed in parallel, and then the results of both are added with the residual and passed to the next layer as shown in Equation~\ref{eq:ffn2}. This architecture can improve the computing throughput by about 15\%.
\begin{eqnarray}
\small
    x_{t+1}&=&x_{t}+FFN(LN(x_t+Attn(LN(x_t)))) \label{eq:ffn1}\\
    x_{t+1}&=&x_{t}+Attn(LN(x_t))+FFN(LN(x_t)) \label{eq:ffn2}
\end{eqnarray}

\textbf{Activations.}  We use \tu{GeLU}~\cite{hendrycks2023gaussian} activation function, which can alleviate the issue of gradient vanishing during the training process. Moreover, it introduces a nonlinear transformation similar to the \tu{sigmoid} function, which helps to accelerate the convergence speed of the model. 

\textbf{Layer normalizations.} We adopt the pre-layer normalization~\cite{ba2016layer} to the input of the Transformer block which is more robust to input variations and improved gradient flow~\cite{xiong2020layer}. 

\subsection{Pre-training}

We trained \codefuse based on the GPT-NeoX~\cite{black2022gptneox20b} framework, which is built on Megatron~\cite{shoeybi2020megatronlm} and DeepSpeed and incorporates deep optimizations for algorithms, communication, and model parallelism. We trained \codefuse on a 64-node GPU cluster, each with eight NVIDIA A100-SXM-80GB GPUs. We trained \codefuse in various sizes with 350M, 1.3B, 6B, and 13B parameters, details as shown in Table~\ref{tab:sst}.

\begin{table*}[t]
  \caption{Family models of \codefuse.}
  \label{tab:sst}
  \begin{tabular}{cccccccc}
    \toprule
    \textbf{Model} &\textbf{NumLayers}& \textbf{NumHeads} & \textbf{HiddenSize}& \textbf{SeqLen} & \textbf{BatchSize}& \textbf{LearningRate} & \textbf{Paralells}\\
    \midrule
   \codefuse-350M	& 24& 	16&	1024&	2048&	1024&	2e-4&	DP=64\\
\codefuse-1.3B&	24&	16&	2048&	2048&	1024&	2e-4&	DP=128\\
\codefuse-6B&	28&	32&	4096&	4096&	2048&	1.5e-4&	DP=256\\
\codefuse-13B&	40&	40&	5120&	4096&	4096&	1.5e-4&	DP=256, TP=2 \\
  \bottomrule
\end{tabular}
\end{table*}

\textbf{Training optimizations.} \codefuse-13B is trained with 256-way data parallelism, 2-way tensor parallelism, and sequence parallelism, and reducing memory consumption with DeepSpeed ZeRO-1~\cite{rajbhandari2020zero}. The sequence length of \codefuse-13B is 4096, and accelerate long sequence model training with Flash Attention~\cite{dao2022flashattention}. The micro batch size is 16,  and the global batch size is 4096, we achieved 180 TFLOPS and 56\% average utilization rate of tensor cores on 512 GPUs. 
We use Adam~\cite{kingma2017adam} optimizer for training, where the initial learning rate is 1.5e-4, and the min learning rate is 1.5e-5, along with cosine learning rate decay style. 
We use fp16 mixed precision training mode. To avoid precision underflow or overflow, we set the initial loss scaling to 32768 and the minimum loss scaling to 1. 


        

\subsection{Supervised Finetuning}
The supervised finetuning (SFT) of \codefuse containts several dimensions: data collection, instruction augmentation, data formatting, training strategy, and fine-tuning framework. The following sections will elaborate on each dimension.

\textbf{Data collection for SFT. }
As a domain-specific LLM for the code field, it needs to have the following functionalities:
\begin{itemize}
    \item Basic natural language understanding capability: The model should provide a satisfactory understanding of complex natural language questions for text2code tasks.
\item Common downstream code tasks: Tasks such as Text2Code (natural language to code snippet), CodeTrans(code translation between programming languages), CodeComment (code commenting), CodeExplain (code explanation), TestCase (generating test cases), and more.
\item Multi-turn dialogue capability: It should support multi-turn conversations where user queries may or may not be related to the previous context, requiring the model to make accurate intent judgments.
\item Non-toxic output: The model should generate content that is free from toxicity and harm. Besides adhering to basic human values and moral ethics, \codefuse, being a code LLM, needs to pay special attention to code-related toxicity. For instance, it should not output content that may pose information security issues, such as fishing or Trojan programs.
\end{itemize}

\textbf{Self-instruct instruction augmentation.}
We collected question-answer pairs for code-related tasks from the public domain as well as code semantic extraction as described in 
Section~\ref{sec:semantic}. 
However, the dataset for code tasks from public domain is much smaller than that of language tasks, and the code semantic extraction is restricted for certain code tasks. 
This requires new ways to augment datasets for poverty tasks. 
Thanks to the self-instruct techniques introduced in Alpaca~\cite{alpaca}, \codefuse leverages those high-quality open or manual writing original dataset as seeds, and generate informative and contextually relevant outputs with the help of off-the-shelf models like ChatGPT. This approach is suitable for divergent data augmentation.

\textbf{Instruction-following description.} The mentioned methods allow us to construct suitable inputs and outputs for different tasks. However, it still has a step gap for an LLM to follow instructions, that is instruction description. 
Some data may already have instructions included in the input, while others require adding instruction descriptions to familiarize the model with the task-specific instructions. This enables the model to be triggered to follow similar instruction descriptions during usage. Moreover, in-context description and chain-of-thoughts (CoT) information can be added to enhance the model's performance.

\textbf{Data format.} In practice, data for different scenarios, such as multi-turn conversations, few-shot learning, and CoT tasks, comes from diverse channels and has complex formats. To handle this, we standardized the original JSON format for these different task types. 
Taking into account existing practices in academia and industry, we have ultimately decided to adopt ChatML (Chat Markup Language)~\cite{chatml} and have made optimizations accordingly. We have proactively addressed issues such as tokenization errors, instruction injection attacks, and Multi-Role-Playing scenarios in the format design. 
LLM model accepts a sequence of text inputs during training and inference and converts the original JSON data into a sequence of text in ChatML format. 
\codefuse dialogue model supports three roles by default: System, Human, and Bot. The System role provides initial information and instructions, while the Human and Bot roles represent user inputs and model-generated responses, respectively.

\textbf{Optimizations.}  During fine-tuning, the model focuses on learning the Bot's output by calculating the loss only for the Bot role. Additionally, \codefuse introduces multi-task finetuning (MFT) since downstream code tasks are naturally divided into multi-tasks. MFT introduces well-designed loss functions based on multi-task learning that enables the model to effectively learn from each task even with different sample numbers, difficulties, and convergence speed.
MFT can complement different tasks to achieve better results than SFT. Details on MFT will be elaborated in another paper.

\begin{figure}[]
\lstset{frame=lrtb,
  aboveskip=3mm,
  belowskip=3mm,
  xleftmargin=2mm,
  xrightmargin=2mm,
  showstringspaces=false,
  columns=flexible,
  basicstyle={\small},
  numberstyle=\tiny\color{gray},
  keywordstyle=\color{blue},
  commentstyle=\color{dkgreen},
  stringstyle=\color{mauve},
  breaklines=true,
  breakatwhitespace=true,
  tabsize=2,
  escapeinside={(*@}{@*)},
}
\begin{lstlisting}
<|im_start|>system 
Provide some context and/or instructions to the model.
<|im_end|> 
<|im_start|>user 
The user's message goes here
<|im_end|> 
<|im_start|>assistant
\end{lstlisting}
\caption{An example of ChatML.}
\end{figure}

\subsection{Model operations}

\codefuse-13B was trained using 512 Nvidia A100 GPU cards, with a Hardware FLOPs Utilization (HFU) of approximately 60\%. The training process took approximately 40 days to complete. Several key stability-related capabilities were developed to ensure the successful training process. In particular, we developed a cloud-based observability system with two major parts.
\begin{itemize}
    \item \textbf{Training metrics observability} is essential for monitoring the training process, including important metrics like training/validation loss to assess convergence and computational FLOPs levels. \codefuse has deployed a TensorBoard instance in the cloud, allowing users to access and analyze these metrics through a web browser.
    \item \textbf{Infrastructure metrics observability} covers various aspects such as GPU and RDMA. \codefuse uses DCGM/NVML to gather GPU/RDMA performance metrics, ECC errors, and Xid errors. They also employ a node-side detection agent named Walle, to capture hardware anomaly information. The GPU Diagnose component handles fault recovery operations. The GPU cluster utilizes RDMA high-performance networking technology for efficient data transmission and processing.
\end{itemize}

We developed a GPU diagnosis system for proactive discovery and automatic handling failures of GPU nodes, as well as automatically restarting/recovering terminated training tasks without manual intervention. Within 30 minutes of a failure, the system identifies, isolates, and reschedules faulty GPU cards, allowing training to continue with the latest checkpoint. To maintain the high availability of GPU resources, we define the SLO metric for the schedulability of GPU cards, resulting in an increase of GPU availability from 87\% to 94\% for the cluster during the training cycle.



\section{Evaluation}\label{sec:eval}


In this section, we begin by introducing the models we assessed alongside \codefuse. Our experiments are conducted by using the NVIDIA A100-SXM-80GB GPUs with a Linux system. We present a comprehensive analysis of the performance of all models on the \humaneval, \humaneval-x~\cite{humaneval} and our \codefuseeval benchmarks. We conduct an evaluation of \codefuse in comparison to \codegeex, utilizing a variety of code tasks with Chinese prompts.


\subsection{Evaluation benchmarks and protocols}

\humaneval, its extension \humaneval-x and \mbpp are widely used benchmarks in the field of code LLMs.
These benchmarks encompass a vast collection of programming problems, employing test cases to validate the code generated by code LLMs. However, when it comes to Chinese inputs, there is still a need for evaluation methods in certain code scenarios. 
To handle this issue, we developed and open-sourced \codefuseeval\footref{foot:codefuseeval} to facilitate evaluation in code completion, code generation, cross-program-language code translation, code commenting, and test case generation scenarios with both English and Chinese prompts. 
The \codefuseeval benchmark is an extension of the \humaneval and \mbpp benchmarks, covering five programming languages: Python, Java, C++, Go, and JavaScript. 

For code generation tasks including code completion, text2code, code translation, and test cases generation, we adopt the \tu{pass@k} metric as the evaluation criterion. 
For code commenting tasks, we use \tu{Bleu} and \tu{BleuRT} as the evaluation metrics. 
To comprehensively assess the model's code capabilities in the multi-task evaluation, three decoding strategies were employed: temperature sampling, greedy, and beam search. The prompts were formatted using zero-shot, allowing the model to generate responses without task-specific fine-tuning. 
When employing the temperature sampling strategy, we set the hyperparameters for \tu{pass@1} as follows: \tu{temperature}=0.2, \tu{top\_p}=0.95, and \tu{generaten} = 10 samples.

\subsection{Compared models}
We compare \codefuse to the following models with similar sizes. The statistic of compared models is from the published reports. 
\begin{itemize}

\item \neox-20B~\cite{black2022gptneox20b} is a 20 billion parameter autoregressive language model trained on the Pile~\cite{gao2020pile}.

\item \starcoder~\cite{starcoder} is a Code LLM with 15B parameters and a context size of 8K, which supports infilling capabilities and fast inference.

\item \codegeex is a language model that has been trained on a collection of 23 programming languages. It is an open-sourced model with 13 billion parameters. Its training data was selected from the Pile\cite{gao2020pile},  CodeParrot \cite{wolf-etal-2020-transformers}, and other datasets. In addition to these datasets, \codegeex also includes its own multi-language benchmark suite, \humaneval-x, which we discuss below.

\item \ernie\cite{sun2019ernie,sun2019ernieb,wang2021ernie} (Enhanced Representation through kNowledge IntEgration) is a language representation model enhanced by using knowledge masking strategies. The masking strategy of \ernie inspired by BERT\cite{devlin2019bert} includes phrase-level strategy and entity-level strategy. 


\item \codegen~\cite{nijkamp2023codegen} has two versions, \codegen-Mono-16B is a variant of \codegen-Multi-16B, specifically fine-tuned using additional Python code from GitHub.
\item CodeT5+~\cite{wang2023codet5}, an encoder-decoder based Code LLM, boasts modular flexibility, accommodating diverse code-related downstream tasks.

\item WizardCoder~\cite{luo2023wizardcoder} is trained using the Evol-Instruct technique. It has shown remarkable improvement in performance on \humaneval python evaluations compared to previous models.

\item PanGu-Coder2~\cite{shen2023pangucoder2} is trained by RRTF (Rank Responses to align Test\&Teacher Feedback) framework with the same Evol-Instruct technique as WizardCoder. PanGu-Coder2 has achieved the leading performance on \humaneval among models of similar size. Like WizardCoder, PanGu-Coder2 is also a mono-lingual model. 

\item CodeLlama~\cite{codellama} is a family of large language models for code, based on Llama 2. It stands out among other open models with its state-of-the-art performance, infilling capabilities, support for large input contexts, and zero-shot instruction following ability for programming tasks.


\end{itemize}

\subsection{Evaluation on code generation}
\begin{table*}[]
  \caption{
  Performance comparison of \codefuse with previous models with similar size on \humaneval-x.
}
\vspace{-1ex}
  \label{tab:eval}
  \begin{tabular}{clccccc}
    \toprule
    \multirow{2}{*}{\textbf{Lingual}}& \multirow{2}{*}{\textbf{Models}}  & \multicolumn{5}{c}{\textbf{\humaneval-x pass@1}}  \\ \cline{3-7}
    &  & \textbf{Python}& \textbf{Java} & \textbf{C++} & \textbf{JavaScript} & \textbf{Go}  \\
    \midrule
\multirow{2}{*}{\textbf{\codefuse}}&\codefuse-13B-Base&	24.83\%&	23.78\%&	22.08\%&	19.62\%&	18.17\%\\
&\codefuse-13B-SFT&			37.10\%&	26.22\%&	19.51\%&	31.71\%&	24.39\%\\
\midrule
\multirow{7}{*}{\textbf{Multi-}}&GPT-NeoX-20B&			13.83\%&	8.87\%&	9.90\%&	11.28\%&	5.00\%\\
&CodeGEEX-13B&			22.89\%&	20.04\%&	17.06\%&	17.59\%&	14.43\%\\
&Baidu-ERNIE-3.5-15.5B&		35.37\%&	26.22\%&	20.11\%&	34.76\%&	27.43\%\\
&StarCoder-15.5B&		33.57\%&	30.22\%&	31.55\%&	30.79\%&	17.61\%\\
&\codegen-multi-16B&		19.22\%& 14.95\%&18.05\%&18.40\%&13.03\%\\
&CodeT5+-16B& 		30.90\%&	&&&\\
&CodeLlama-13B&	36.00\%&	&&&\\
\midrule
\multirow{3}{*}{\textbf{Mono-}}&CodeLlama-Python-13B& 	43.30\%&	&&&\\
&WizardCoder-16B& 	57.30\%&	&&&\\
&PanGu-Coder2-15B& 	61.64\%&	&&&\\

\bottomrule
\end{tabular}
\end{table*}

In our comparison of \codefuse with existing Code Language Models (LLMs) of similar size, we evaluated their code generation performance. To ensure fairness, we gathered statistics for other models in Table~\ref{tab:eval} from existing reports. However, the performance of CodeT5+ and CodeLlama on \humaneval-x was not presented in their respective articles or other published related work. As a result, Table~\ref{tab:eval} lacks data for these models.

We present the performance of multiple versions of \codefuse because they demonstrate the progress of \codefuse and achievements in different tasks. Most of these versions are deployed in various scenarios.
Additionally, we showcase seven current mainstream multi-lingual models ranging in size from 13 billion to 16 billion parameters. We also present three state-of-the-art code Language Model Models (LLMs), among which PanGu-Coder2, released last month, achieved the highest \humaneval score.

In the \humaneval Python pass@1 evaluation, the open-sourced version of \codefuse-13B outperforms other multi-lingual models. However, since mono-lingual models like CodeLlama-Python, WizardCoder, and PanGu-Coder2 have made specific optimizations for Python code generation, it is still challenging for \codefuse to surpass them in Python evaluation. 




\subsection{Evaluation on multi-lingual code translation}
We evaluate \codefuse on multi-lingual code translation by comparing it to \codegen-multi-16B and \codegeex-13B. 
Similarly to \codegeex-13B, which has a dedicated fine-tuned version for code translation called \codegeex-13B-FT, \codefuse-13B-SFT is a multi-task fine-tuned version that includes code translation. 

The code translation evaluation dataset we used is constructed based on \mbxp and \humaneval-x. It includes cases that have been reviewed and corrected by experts and has been opened as a part of \codefuseeval. 

Table~\ref{tab:codetrans} presents the pass@1 results for mutual conversion between Java, Python, and C++ using greedy decoding. The table clearly shows that \codefuse-13B-SFT outperforms other models when translating Python code to the other two languages. Additionally, it achieves the highest average score across all six translation scenarios. 

\begin{table*}[ht]
  \caption{Performance(pass@1) comparison of \codefuse with previous models on code translation using greedy decoding}
\vspace{-1ex}
  \label{tab:codetrans}
  \begin{tabular}{lccccccc}
    \toprule
     {\textbf{Models}} & \textbf{Java to Py}&	\textbf{C++ to Py}&	\textbf{C++ to Java}&	\textbf{Java to C++}&	\textbf{Py to Java}&	\textbf{Py to C++} & \textbf{Average}\\
     \midrule
\codefuse-13B-Base&	53.66\%	&	55.49\%	&	41.46\%	&	37.80\%	&	48.10\%	&	50.00\%	& 47.75\%\\
\codefuse-13B-SFT&	66.46\%	&	59.15\%	&	54.27\%	&	47.56\%	&	\textbf{56.31\%}	&	\textbf{55.40\%}&	\textbf{56.53\%}\\
\codegen-multi-16B& 52.73\%	&	33.83\%	&	43.20\%	&	41.42\%	&	29.27\%	&	35.94\%	& 39.40\%\\
\codegeex-13B&	43.41\%	&	27.18\%	&	22.56\%	&	39.33\%	&	25.84\%	&	26.54\%	& 30.81\%\\
\codegeex-13B-FT&	\textbf{75.03\%}	&	\textbf{62.79\%}	&	\textbf{71.68\%}	&	\textbf{49.67\%}	&	41.98\%	&	34.16\%	& 55.89\%\\
\bottomrule
\end{tabular}
\end{table*}

\subsection{Evaluation on code-related tasks with Chinese prompts}

\begin{figure*}
\lstset{frame=lrtb,
  language=python,
  aboveskip=3mm,
  belowskip=3mm,
  xleftmargin=2mm,
  xrightmargin=2mm,
  showstringspaces=false,
  columns=flexible,
  basicstyle={\small},
  numberstyle=\tiny\color{gray},
  keywordstyle=\color{blue},
  stringstyle=\color{mauve},
  breaklines=true,
  breakatwhitespace=true,
  tabsize=2,
  escapeinside={(*@}{@*)},
}

\begin{subfigure}{0.45\textwidth}
\includegraphics[width=\textwidth]{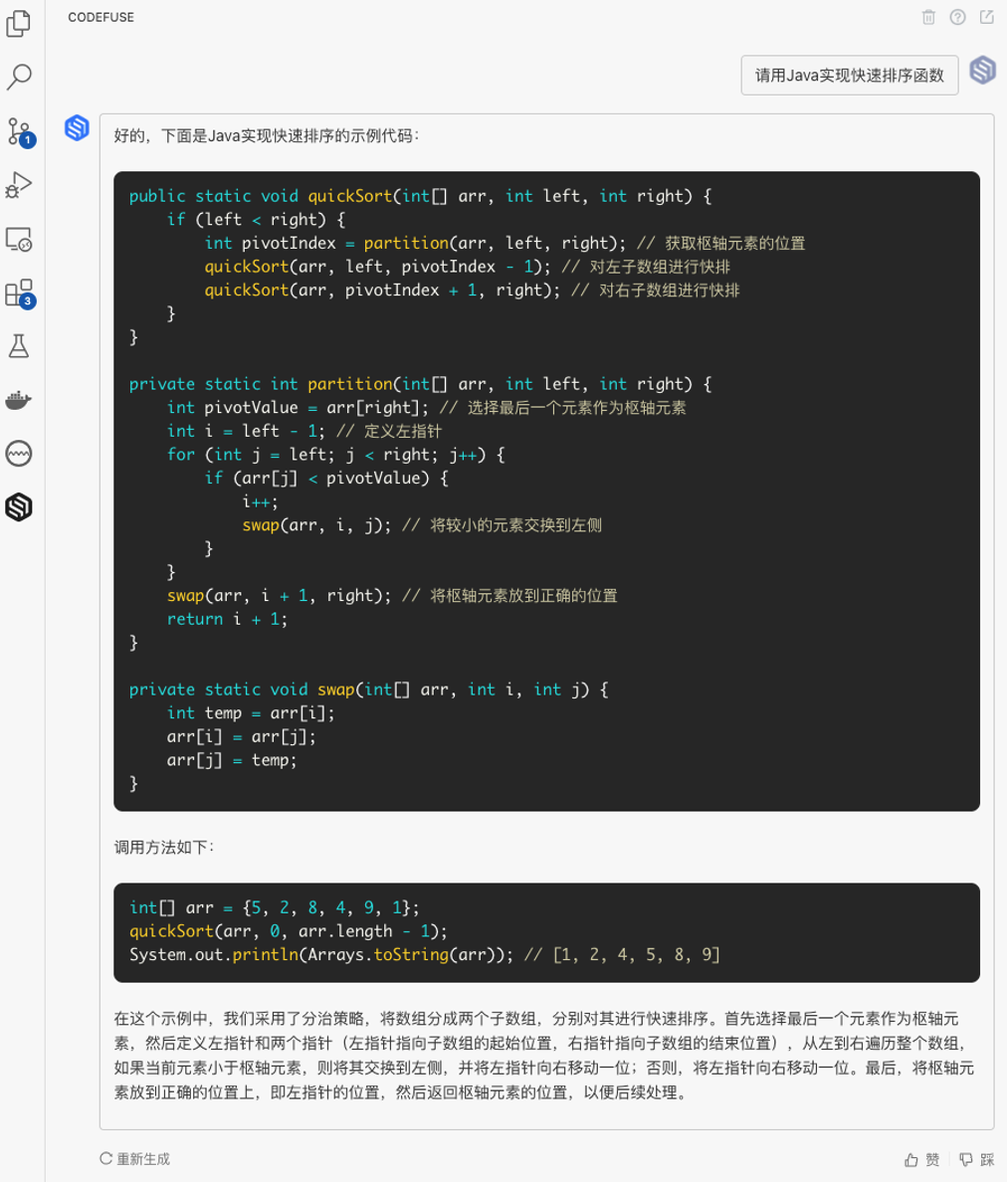}
\subcaption*{a. An example of code generation.}
\label{fig:cn_codegen_example}
\includegraphics[width=\textwidth]{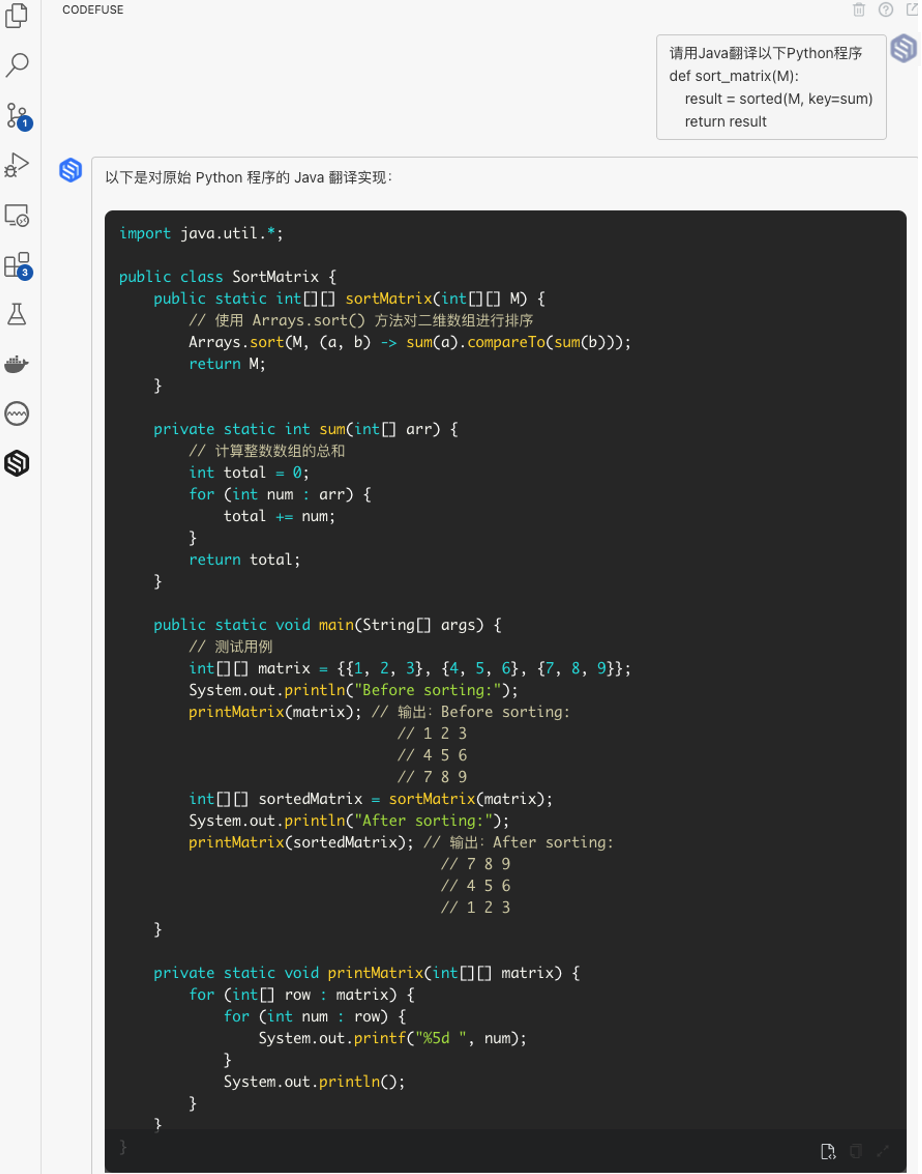}
\subcaption*{b. An example of code translation.}
\label{fig:cn_codetrans_example}
\end{subfigure}
\begin{subfigure}{0.42\textwidth}
\includegraphics[width=\textwidth]{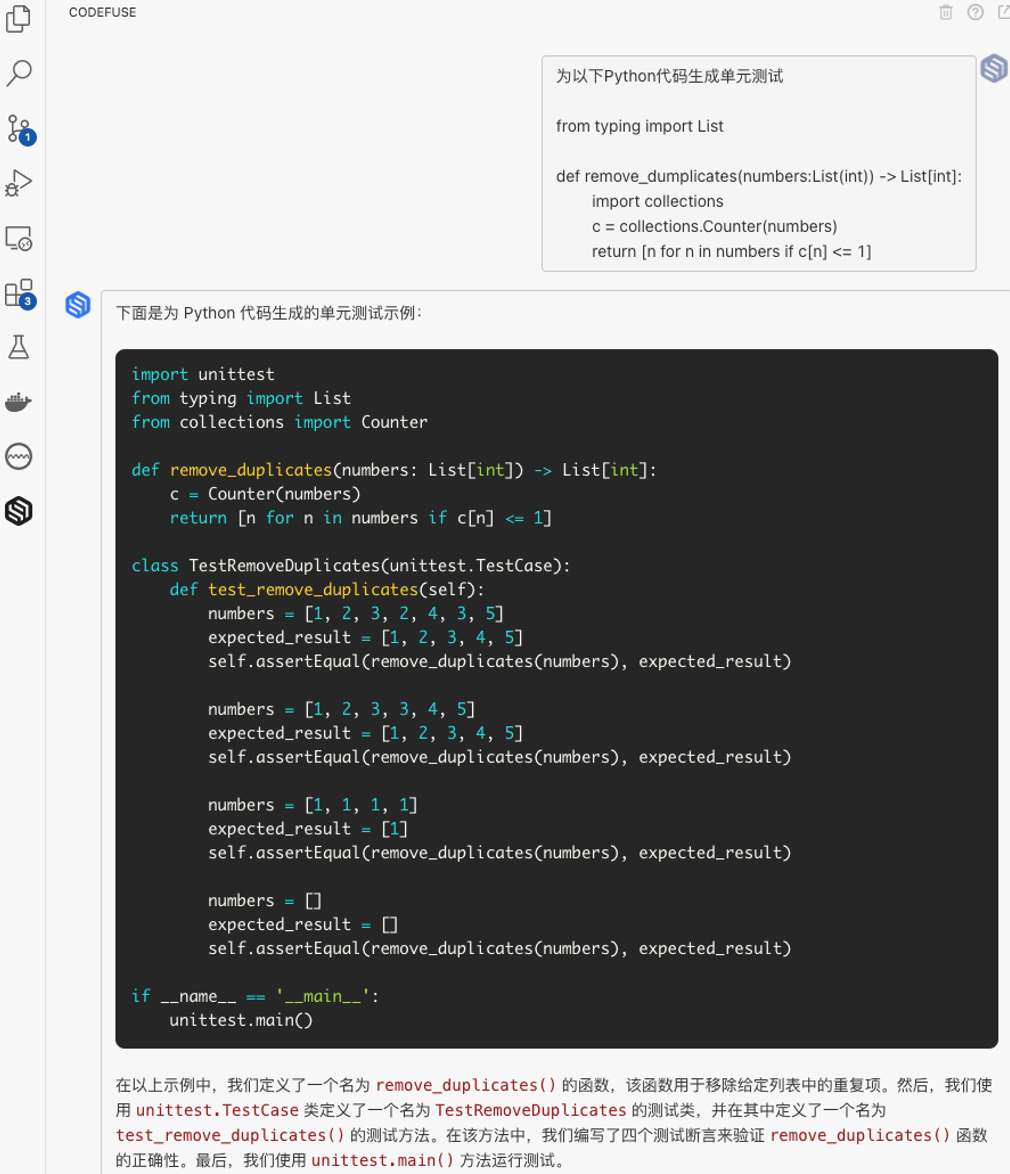}
\subcaption*{c. An example of testcase generation.}
\label{fig:cn_testcase_example}
\includegraphics[width=\textwidth]{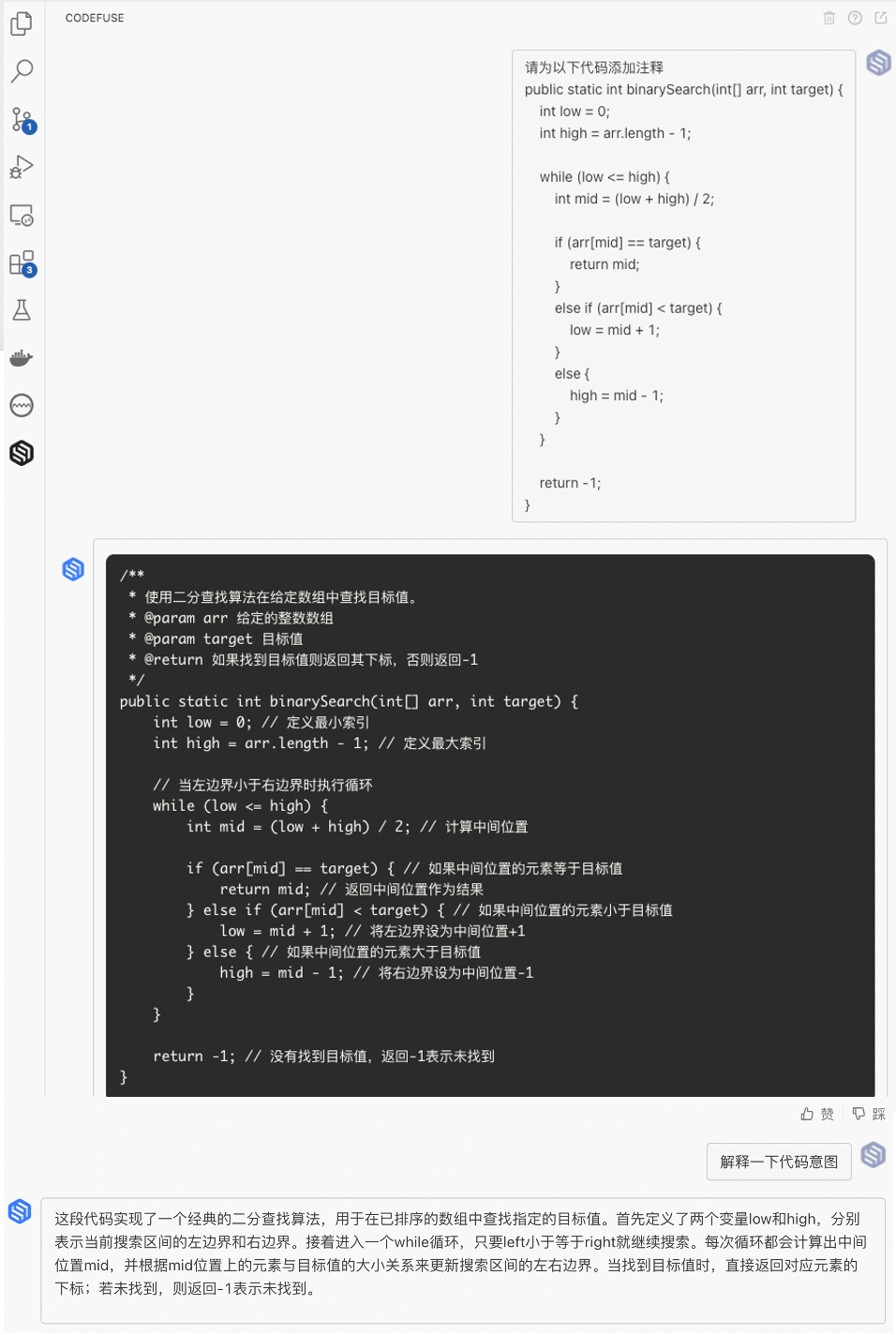}
\subcaption*{d. An example of code comments and explanation.}
\label{fig:cn_comments_example}
\end{subfigure}

\caption{Examples of \codefuse generation with prompts in Chinese. The results are generated by \codefuse VSCode extension, and indicators such as ``\#'' have been removed for better human readability.
\label{fig:cn_examples}}
\end{figure*}

We conducted an evaluation of \codefuse on supporting Chinese prompts by comparing it to \codegeex, which is known for its excellent Chinese language support. Figure~\ref{fig:cn_examples} showcases examples of code generation, code translation, code comments, and test case generation with prompts in Chinese. In addition to the evaluation of code generation and translation mentioned in the previous sections, testcase generation, code comments, and explanation are included in our evaluation.

In the code comment and explanation tasks, We specifically evaluated our \codefuse, by integrating it into \company's development process. Valuable feedback from developers during daily work was collected to assess performance in real-world scenarios. We did not compare \codefuse with other models due to the time and effort required to deploy a competing model in our daily development process.
After collecting human feedback over several months, \codefuse-13B-SFT achieved a \tu{Bleu} score of 42.42\% and a \tu{BleuRT} score of 36.34\% as shown in Table~\ref{tab:chcc}. These results indicate that \codefuse is indeed useful in real development scenarios.

\begin{table}[h]
  \caption{Evaluation of Chinese code comments and explanations on \codefuseeval and real feedback}
  \label{tab:chcc}
  \begin{tabular}{lcccc}
    \toprule
     {\textbf{Models}} & {\textbf{Bleu}}& {\textbf{BleuRT}}  \\ 
     
    \midrule
\codefuse-13B-Base &	36.75\%&	27.76\%\\
\codefuse-13B-SFT&	42.42\%&	36.34\%\\
\bottomrule
\end{tabular}
\end{table}

Recently, there have been efforts to automatically generate test cases by providing a code snippet to LLMs~\cite{schafer2023empirical,wang2023software,10.1145/3453483.3454054}. In our evaluation, we used the \humaneval-x dataset and selected \codegeex as the baseline model due to its comparable model size, support for Chinese prompts, and relevance to industry applications. We made minor adjustments to the dataset to adhere to the prompt format shown in Figure~\ref{fig:cn_examples}(c). As shown in Table~\ref{tab:chtg}, \codefuse outperforms \codegeex in the task of testcase generation with Chinese prompts.

\begin{table}[h]
  \caption{Evaluation of Chinese testcase generation on \codefuseeval}
  \label{tab:chtg}
  \begin{tabular}{lcc}
    \toprule
     {\textbf{Models}} & {\textbf{pass@1 Python}}& {\textbf{pass@1 Java}}  \\ 
    \midrule
\codefuse-13B-SFT	&31.20\%	&24.32\% \\
\codegeex-13B	&22.89\%	&20.04\% \\
\bottomrule
\end{tabular}
\end{table}

\section{Related work}

\textbf{Large language models. }
LLMs have exhibited remarkable accomplishments across a wide range of tasks. Leading technology companies have made substantial progress in creating highly capable LLMs. Notable examples include OpenAI's GPT3\&4~\cite{brown2020language,openai2023gpt4}, Google's PaLM~\cite{chowdhery2022palm,anil2023palm}, DeepMind's Chinchilla~\cite{hoffmann2022training} and Gopher~\cite{rae2022scaling}, as well as Anthropic's Claude4. However, these models are closed-source and only accessed through specific APIs.


Several open-source LLMs have been released and made valuable contributions to the public. EleutherAI has contributed \neox-20B~\cite{black2022gptneox20b} and GPT-J-6B~\cite{gpt-j}. Google has released UL2-20B. Tsinghua University has introduced GLM-130B~\cite{codegeex,zeng2022glm130b}. Meta has released LLaMA~\cite{llama} and LLaMA2~\cite{llama2}.

\textbf{Code large language models. }
Many works have introduced LLMs to tackle the challenges of code understanding and generation problems. Codex~\cite{chen2021evaluating} has powered Copilot for code tasks. Google has proposed PaLM-Coder~\cite{chowdhery2022palm}. These models have shown exceptional performance on popular code completion benchmarks such as \humaneval~\cite{humaneval} and \mbpp~\cite{austin2021program}. However, it is important to note that these models are closed-source.

There also are several open-source Code LLMs available. Salesforce has developed CodeGen~\cite{nijkamp2023codegen}, CodeT5~\cite{wang2021codet5}, and CodeT5+~\cite{wang2023codet5}. Tsinghua University has contributed CodeGeeX~\cite{codegeex}, and the BigCode Project has created StarCoder~\cite{starcoder}. While recent models like WizardCode~\cite{luo2023wizardcoder}, PanGu-Coder2~\cite{shen2023pangucoder2}, and CodeLLaMA~\cite{codellama} have achieved impressive scores on \humaneval, it is worth noting that they are weak for multi-lingual prompts. In contrast, \codefuse is designed to be a multi-lingual code Language Model (LLM) that supports various code-related tasks across both English and Chinese prompots. This makes \codefuse a valuable solution for developers working in diverse linguistic environments.

\section{Discussion, Conclusion, Future Work}

\textbf{Extending MFT framework to support opensource models.}
We develop a multi-task finetuning framework and open-source it as MFTCoder\footnote{\url{https://github.com/codefuse-ai/MFTCoder}}.
This framework can be used to fine-tune both our in-house developed \codefuse-13B and newly emerging open-sourced Code LLMs such as StarCoder~\cite{starcoder} and CodeLLaMA~\cite{codellama}. 
We fine-tuned StarCoder and CodeLLaMA
models with the MFTCoder framework on our collected datasets, and open-sourced the fine-tuned version as \codefusellama-34B
and \codefuse-\starcoder-15B. 
\codefusellama-34B achieves 74.4\% pass@1 score on \humaneval, which surpasses the score by GPT4 and ChatGPT-3.5, and represents the state-of-the-art results for open-sourced Language Model Models (LLMs). It is also evident that the data preparation strategy in \codefuse greatly enhances the performance of other code LLMs.

\textbf{Deployment.} 
\codefuse is deployed in the production environment within \company in the form of both IDE plugin and web-based chat. See Figure~\ref{fig:cn_examples} for some examples. To facilitate the model response time and service throughput, we introduce a series of optimizations for the model service, which include (1) quantizing the model to 4bits with negligible accuracy loss using automatic iterative refined GPTQ~\cite{gptq}; (2) leveraging software optimization provided by Nvidia TensorRT-LLM~\footnote{\url{https://developer.nvidia.com/tensorrt-llm-early-access}}; (3) performing service optimization through semantic cache and streaming output. The product now supports daily software developing of more than 5K engineers in \company.   

\textbf{Conclusion.}
This paper introduces \codefuse-13B, an open-sourced pre-trained Language Model (LLM) with 13 billion parameters designed for code-related tasks with multi-lingual (English and Chinese) prompts. It supports over 40 programming languages and utilizes a carefully filtered pre-training dataset. Experiments using real-world scenarios and industry benchmarks demonstrate that \codefuse-13B achieves a \humaneval Pass@1 score of 37.10\%, making it one of the top multi-lingual models with similar parameter sizes. It outperforms other models in code generation, translation, comments, and testcase generation tasks with Chinese inputs. Valuable human feedback from \company's software development process confirms the successful integration of \codefuse-13B.




\textbf{Future work. }
Besides models of \codefuse and the MFTCoder framework, we plan to further open-source two significant components of \codefuse: the \codefuseeval benchmark and the \tu{Sparrow} program query system for high-quality code data cleaning. 
By open-sourcing these components, we aim to contribute to the research community and facilitate further advancements in the full lifecycle of AI native software development.





\bibliographystyle{ACM-Reference-Format}
\bibliography{ref}




\end{document}